\documentclass[]{elsart}

\usepackage{graphicx}
\usepackage{amsmath}
\usepackage{amssymb}
\usepackage{bm}

\usepackage{color}
\usepackage{colortbl}

\providecommand{\Journal}[4] {#1 {\bf #2} (#4) #3}
\providecommand{\EPJC}{Eur. Phys. J. C } %
\providecommand{\MPLA}{Mod. Phys. Lett. A} %
\providecommand{\NPB}{Nucl. Phys. B } %
\providecommand{\PA}{Physica A } %
\providecommand{\PLB}{Phys. Lett. B } %
\providecommand{\PLA}{Phys. Lett. A } %
\providecommand{\PR}{Phys. Rep. } %
\providecommand{\PRV}{Phys. Rev. } %
\providecommand{\PRL}{Phys. Rev. Lett. } %
\providecommand{\PRD}{Phys. Rev. D } %
\providecommand{\ZP}{Z. Phys. } %
\providecommand{\ZPC}{Z. Phys. C } %
\providecommand{\IJMPA}{Int. J. Mod. Phys. A } %
\providecommand{\JHEP}{JHEP } %
\providecommand{\JPG}{J. Phys. G } %
\providecommand{\JRE}{Jahrbuch der Radioaktivitat und Electronik } %

\begin{document}

\begin{frontmatter}


\journal{Physics Letters B}

\title{Statistical effect in the parton distribution functions
of the nucleon}

\author{Yunhua Zhang},
\author{Lijing Shao},
\author{Bo-Qiang Ma}\ead{mabq@phy.pku.edu.cn}
\address{School of Physics and State Key Laboratory of Nuclear
Physics and Technology, Peking University, Beijing 100871, China}

\begin{abstract}
A new and simple statistical approach is performed to calculate the
parton distribution functions (PDFs) of the nucleon in terms of
light-front kinematic variables. We do not put in any extra
arbitrary parameter or corrected term by hand, which guarantees the
stringency of our approach. Analytic expressions of the
$x$-dependent PDFs are obtained in the whole $x$ region [0,1], and
some features, especially the low-$x$ rise, are more agreeable with
experimental data than those in some previous instant-form
statistical models in the infinite-momentum frame (IMF). Discussions
on heavy-flavored PDFs are also presented.
\end{abstract}


\end{frontmatter}

\newpage

\section{Introduction}

One of the goals in fundamental physics is to search for the detail
information of the nucleon structure. The parton model, suggested by
Feynman~\cite{f69}, and then immediately applied to the deep
inelastic lepton-nucleon scattering process by Bjorken~\cite{bp69},
proposes that the nucleon is composed of a number of point-like
constituents, named ``partons'', which were afterward recognized as
quarks and gluons. In the impulse approximation, the deep inelastic
lepton-nucleon scattering can be viewed as a sum of elastic
lepton-parton scattering, in which the incident lepton is scattered
off a parton instantaneously and incoherently. This is in accord
with one property of QCD -- asymptotic freedom. On the other hand,
due to another property of QCD -- color-confinement, the
constituents of nucleon -- quarks and gluons, have never been seen
individually. The nucleon structure functions, in terms of the
parton distribution functions (PDFs), are badly desired in hadronic
study. However, due to the complicated non-perturbative effect, we
still have difficulty to calculate them absolutely from the first
principal theory of QCD at present.

Various models according to the spirit of QCD have been brought
forward, therein statistical ones, providing intuitive appeal and
physical simplicity, have made amazing
success~\cite{c87,ct88,mu89,bl90,cdj93,gdr91,dkg94,dm96,bh95,bh96,bkr00,bh00,bs95,bsb02,bsb05,zzm01,zzy02,zdm02,ah05}.
Actually, as can be speculated, with partons bound in the wee volume
of the nucleon, not only the dynamic, but also the statistical
properties, for example, the Pauli exclusion principle, should have
important effect on the PDFs. Cleymans and Thews~\cite{c87,ct88}, as
pioneers, started with the transition rate of scattering in the
framework of temperature dependent field theory and explored a
statistical way to generate compatible PDFs. Mac and
Ugaz~\cite{mu89} incorporated first order QCD corrections (however,
the perturbative term turned out to be a sizable fraction of the
statistical term), and afterwards Bhalerao {\it et
al.}~\cite{bh95,bh96,bkr00,bh00} introduced finite-size correction
and got more fitting results; they both referred to the
infinite-momentum frame (IMF). Bickerstaff and Londergan~\cite{bl90}
interpreted the finite-temperature property as to mimic some of the
volume-dependent effects due to confinement, furthermore, they
discussed the theoretical validity of the ideal gas assumption in
detail. Devanathan {\it et al.}~\cite{gdr91,dkg94,dm96} proposed a
thermodynamical bag model which evolves as a function of $x$, and
the structure functions they got are practicable for $x>0.2$ and
have correct asymptotic behavior as $x\rightarrow 1$; in addition,
they parametrized on $T$ and exhibited the scaling behavior.
Bourrely, Soffer and Buccella~\cite{bs95,bsb02,bsb05} developed a
new form of statistical parametrization, allowing $x$-dependent
chemical potential, and by incorporating QCD evolution they got
indeed remarkable PDFs. Otherwise, Zhang {\it et
al.}~\cite{zzm01,zzy02,zdm02} constructed a model using the
principle of detailed balance and balance without any free
parameter, and the Gottfried sum they got is in surprisingly
agreement with experiments. Alberg and Henley~\cite{ah05} tracked
the detailed balance model for a hadron composed of quark and gluon
Fock states and obtained parton distributions for the proton as well
as the pion.

When dealing with physics in high energy region, light-front
dynamics is a suitable language~\cite{bpp98}. It is
known~\cite{ms90} that the impulse approximation fails when using
instant-form dynamics in the nucleon rest frame, but works well when
using it in the IMF or using light-front dynamics in an ordinary
frame. However, statistics in light-front formalism encounters some
difficulties, therein the most fatal one is the unclarity about what
the light-front temperature is and how it relates with the usual one
in instant-form formalism (see, {\it e.g.},
Refs.~\cite{adp02,w03,bmfw03,rb04}). Consequently, the way of
generalization from instant-form statistics to light-front
statistics is quite speculative. Actually, even the generalization
of the thermal dynamics and statistical theory from the system rest
frame to a moving frame in instant-form formalism is also not so
understood, and discussion on it has continued for a long period
(For theoretical discussion, see
Refs.~\cite{e1907,p58,o63,hw71,cm95,lm04,r0801} and references
therein, and for recent numerical experiments, see, {\it e.g.},
Refs.~\cite{cpdth07,r0804}).

In order to avoid these tough problems, we start with instant-form
statistical expressions in the nucleon rest frame, then perform
transformation on them in terms of light-front kinematic variables.
The analytic expressions of the PDFs we get are something different
from those attained in other statistical models performing in the
IMF~\cite{mu89,bh95,bh96,bkr00,bh00}. The largest distinction is
that, when $x\rightarrow0$, the distributions of our light-flavored
quarks (anti-quarks) do not tend to zero as theirs, but give a rise
instead, which agrees with the experimental data better.

Worthy to note that, our intention is only to illustrate whether the
statistical effect is important and to which aspects of the nucleon
structure it is important, not how well it matches the experimental
results, so we do not make any effort to fit the experimental data
intentionally. There is no arbitrary parameter put by hand in our
model, and all parameters are basic statistical quantities. Some of
other statistical models can fit the experimental data better by
introducing many free parameters, however, it weakens the stringency
at a cost. In addition, our results naturally cover the whole $x$
region [0,1], and the features of PDFs and structure functions at
the boundary are of great interest in both theoretical and
experimental study.

The paper is organized as follows. In section 2, a brief description
about the approach used in this paper is introduced, and the
analytic expressions of the PDFs are presented. In section 3,
numerical results and comparisons with experiments and other
theories are illustrated. In section 4, the mass effect of the
partons and the features of heavy-flavored PDFs are discussed. The
last section is a short summary.

\section{The statistical approach}
We assume that the nucleon is a thermal system in equilibrium, made
up of free partons (quarks, anti-quarks and gluons). In the nucleon
rest frame, the mean number of the parton (denoted by $f$) is
\begin{equation} \label{ni}
\bar{N}_f=\int f(k^0)\,\mathrm{d}^3k\;,
\end{equation}
where $f(k^0)$ satisfies the Fermi-Dirac distribution or the
Bose-Einstein distribution
\begin{equation} \label{fi}
f(k^0)=\frac{g_fV}{(2\pi)^3}\frac{1}{e^\frac{k^0-\mu_f}{T}\pm
1}\;,
\end{equation}
with the upper sign for Fermion (quark, anti-quark), and nether sign
for Boson (gluon); $g_f$ is the degree of color-spin degeneracy,
which is 6 for quark (anti-quark) and 16 for gluon; $\mu_f$ is its
chemical potential, and for anti-quark $\mu_{\bar{q}}=-\mu_q$, for
gluon $\mu_g=0$.

Instead of boosting above expressions to the
IMF~\cite{mu89,bh95,bh96,bkr00,bh00}, we transform them in terms of
light-front kinematic variables in the nucleon rest frame. Before
doing this, note that when doing the integration in Eq.~(\ref{ni}),
the on-shell condition $k^0=\sqrt{{\bf k}^2+m_f^2}$ is needed, where
$k^0$, ${\bf k}=(k^1,k^2,k^3)$, $m_f$ are the energy, 3-momentum and
mass of the parton, respectively. Eq.~(\ref{ni}) can be explicitly
reexpressed as
\begin{equation} \label{nc}
\bar{N}_f=\int f(k^0)\delta\left(k^0-\sqrt{(k^3)^2+{\bf
k}_\bot^2+m_f^2}\,\right)\mathrm{d}k^0\mathrm{d}k^3\mathrm{d}^2k_\bot\;.
\end{equation}

We introduce the light-front 4-momentum of the parton
$k=(k^+,k^-,{\bf k}_\bot)$, where $k^+=k^0+k^3$, $k^-=k^0-k^3$,
${\bf k}_\bot=(k^1,k^2)$, and $k^+=P^+ x=Mx$, where $x$ is the
light-front momentum fraction of the nucleon carried by the parton
and $M$ is the mass of the nucleon. Hereby, the $\delta$-function
and the integral in Eq.~(\ref{nc}) turn to
\begin{eqnarray} \label{dt}
\delta\left(k^0-\sqrt{(k^3)^2+{\bf k}_\bot^2+m_f^2}\right)& =&2k^0\theta(k^0)\delta\!\left(k^2-m_f^2\right)\nonumber\\
&=&\left[1+\frac{{\bf
k}_\bot^2+m_f^2}{(Mx)^2}\right]\theta(x)\delta\!\left(k^--\frac{{\bf
k}_\bot^2+m_f^2}{Mx}\right)
\end{eqnarray}
and
\begin{equation} \label{im}
\mathrm{d}k^0\mathrm{d}k^3\mathrm{d}^2k_\bot=\frac{1}{2}M\mathrm{d}k^-\mathrm{d}x\mathrm{d}^2k_\bot\;.
\end{equation}

Then Eq.~(\ref{nc}) becomes
\begin{equation} \label{nf}
\bar{N}_f =\int f(x,{\bf k}_\bot)\,\mathrm{d}^2k_\bot\mathrm{d}x\;,
\end{equation}
where $f(x,{\bf k}_\bot)$ is
\begin{equation} \label{fxk}
f(x,{\bf
k}_\bot)=\frac{g_fMV}{2(2\pi)^3}\frac{1}{e^\frac{\frac{1}{2}\left(Mx+\frac{{\bf
k}_\bot^2+m_f^2}{Mx}\right)-\mu_f}{T}\pm1}\;\left[1+\frac{{\bf
k}_\bot^2+m_f^2}{(Mx)^2}\right]\theta(x)\;.
\end{equation}

On the trivial assumption that ${\bf k}_\bot$ is transversely
isotropic, we can integrate on $|{\bf k}_\bot|$ analytically and get
\begin{eqnarray} \label{fx}
f(x) &=& \pm\frac{g_fMTV}{8\pi^2}\left\{\left(Mx+\frac{m_f^2}{Mx}\right)\,\ln\left[1\pm e^{-\frac{\frac{1}{2}\left(Mx+\frac{m_f^2}{Mx}\right)-\mu_f}{T}}\right]\right.\nonumber\\
&& \left. \qquad\qquad\quad -2T\text{Li}_2\left(\mp
e^{-\frac{\frac{1}{2}\left(Mx+\frac{m_f^2}{Mx}\right)-\mu_f}{T}}\right)\right\}\,,
\end{eqnarray}
as is mentioned above, the upper sign for Fermion and the nether
sign for Boson. $\text{Li}_2(z)$ is the polylogarithm function,
defined as $\text{Li}_2(z)=\sum_{k=1}^\infty z^k/k^2$. Note that the
expressions of the PDFs (Eq.~(\ref{fx})) are different from those
attained in the previous statistical
models~\cite{c87,ct88,mu89,cdj93,bh95,bh96,bkr00,bh00}.

In practice, the PDFs in a certain system should be constrained with
some conditions. For example, in the proton, they are
\begin{equation}
\label{uv} u_V = \int_0^1
[u(x)-{\overline{u}}(x)]\,\mathrm{d}x=2\;,
\end{equation}
\begin{equation}
\label{dv} d_V = \int_0^1
[d(x)-{\overline{d}}(x)]\,\mathrm{d}x=1\;,
\end{equation}
\begin{equation}
\label{qnor} \int_0^1
[u(x)+{\overline{u}}(x)+d(x)+{\overline{d}}(x)+g(x)]x\,\mathrm{d}x=1\;.
\end{equation}

Thus for the proton, we have three equations
(\ref{uv})(\ref{dv})(\ref{qnor}), and four unknown parameters $T$,
$V$, $\mu_u$, $\mu_d$ (The mass of the proton, $M=938.27$ MeV, is
taken as given). So for a given $T$, the rest parameters $V$,
$\mu_u$, $\mu_d$ can be determined by solving the equations.

\section{Results}
We perform our calculation for the proton, therefore the following
results, if not specially stated, are all for proton. However, the
method is absolutely applicable to the neutron. For convenience, we
just consider the $u$, $d$ flavor and gluon, and take $m_f=0$, which
will be showed, in section 4, as a good approximation.

In practice, we adopt a certain value of $T$ at first, then
numerically solve the equations to get $V$, $\mu_u$, $\mu_d$.
Subsequently, the PDFs can be obtained according to Eq.~(\ref{fx}),
as well as the Gottfried sum
\begin{equation}
\label{sg} S_G
=\int_0^1\frac{F_2^p(x)-F_2^n(x)}{x}\,\mathrm{d}x=\frac{1}{3}+\frac{2}{3}\int_0^1\left[\overline{u}(x)-\overline{d}(x)\right]\mathrm{d}x\;.
\end{equation}

We find that at $T_0=47$ MeV, the Gottfried sum $S_G=0.236$, which
agrees well with the experimental result
$S_G=0.235\pm0.026$~\cite{nmc9194}. We conclude that the temperature
of proton is around 47 MeV, and $V_0\approx 1.2\times 10^{-5}$
MeV$^{-3}$, $\mu_u \approx 64$ MeV, $\mu_d \approx 36$ MeV. The
following results are all given at $T_0=47$ MeV.

Taking proton as a perfect sphere, we can calculate its radius $r_0$
from the volume $V_0$. We get $r_0=2.8$ fm, which seems a little
bigger than the practical value, possibly due to the oversimplified
assumption of the uniform distribution of partons. Worthy to mention
that, the $r_0$ we get, together with $T_0$, is close to what Mac
and Ugaz~\cite{mu89} got with the consideration of first-order QCD
corrections, and their fitted values are $T=49$ MeV, $r=2.6$ fm.

\begin{figure}
\begin{center}
\includegraphics[scale=1]{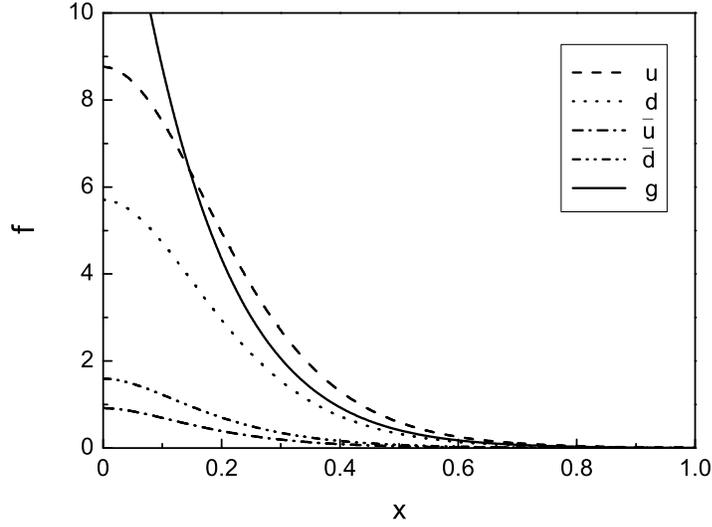}
\caption{The calculated $f(x)$ in our statistical
approach.}\label{f}
\end{center}
\end{figure}

\begin{figure}
\begin{center}
\includegraphics[scale=1]{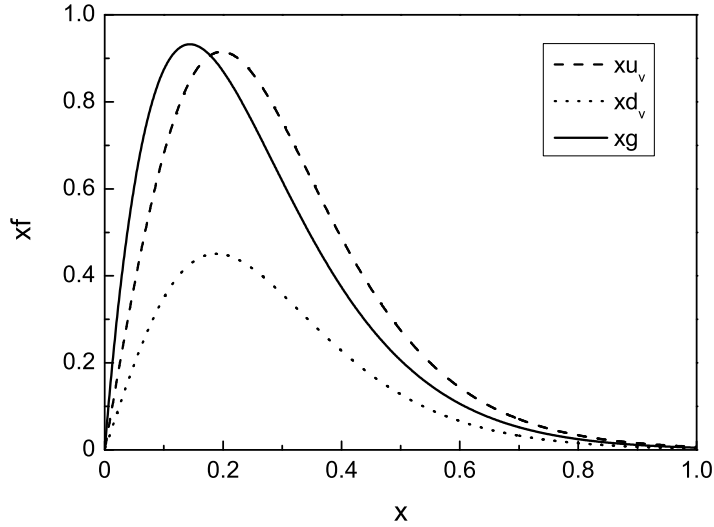}
\caption{The calculated $xf(x)$ in our statistical
approach.}\label{xf}
\end{center}
\end{figure}

The PDFs $f(x)$ and $xf(x)$ are illustrated in Fig.~\ref{f} and
Fig.~\ref{xf}, respectively. In contrast to $q(x)$ and
$\bar{q}(x)\rightarrow0$ as $x\rightarrow0$ in the previous
statistical models without extra corrected
term~\cite{c87,ct88,mu89,cdj93}, our trend makes a good improvement.

\begin{figure}
\begin{center}
\includegraphics[scale=1]{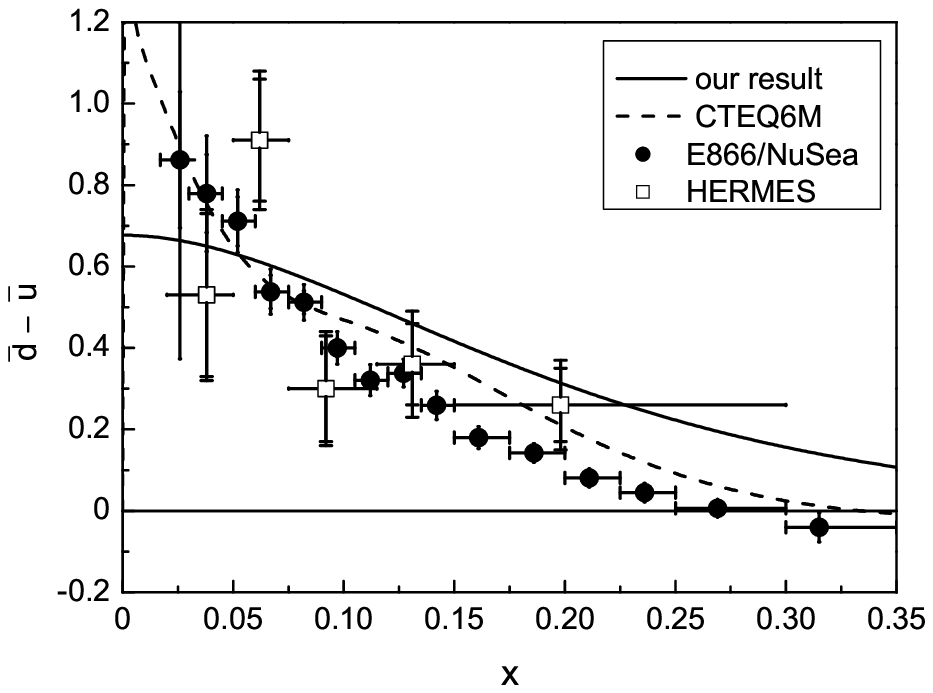}
\caption{Comparison of our result with CTEQ result at $Q^2=1$
GeV$^2$~\cite{cteq02}, E866/NuSea result at $Q^2=54$
GeV$^2$~\cite{e866989901} and HERMES result at $<Q^2>=2.3$
GeV$^2$~\cite{hermes98} for $\bar{d}(x)-\bar{u}(x)$.}\label{dbmub}
\end{center}
\end{figure}

\begin{figure}
\begin{center}
\includegraphics[scale=1]{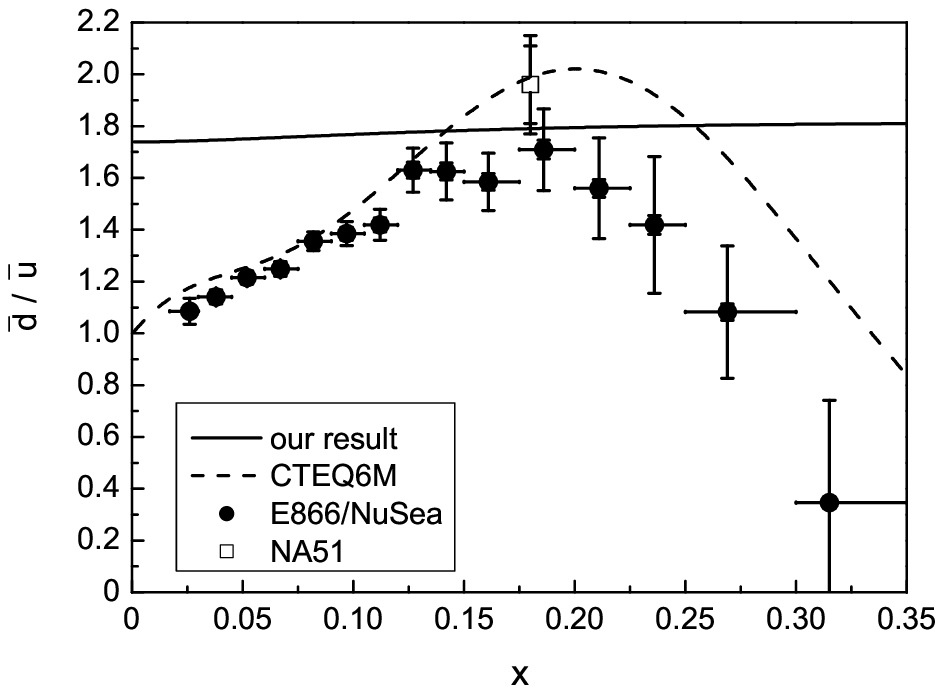}
\caption{Comparison of our result with CTEQ result at $Q^2=1$
GeV$^2$~\cite{cteq02}, E866/NuSea result at $Q^2=54$
GeV$^2$~\cite{e866989901} and NA51 result ~\cite{na94} for
 $\bar{d}(x)/\bar{u}(x)$.}\label{dboub}
\end{center}
\end{figure}

In our model, the flavor asymmetry of the nucleon sea is naturally
generated. $\bar{d}(x)-\bar{u}(x)$ and $\bar{d}(x)\,/\,\bar{u}(x)$
are shown in Fig.~\ref{dbmub} and Fig.~\ref{dboub}, the former
agrees well with the results of experiments and CTEQ
parametrization~\cite{cteq02} while the latter seems not. The model
suggested by Zhang {\it et al.}~\cite{zzm01,zzy02,zdm02} also gives
reasonable asymmetry of $\bar{u}$ and $\bar{d}$ without introducing
any parameter, which is further discussed by Alberg and
Henley~\cite{ah05}. And the feature of $\bar{d}(x)-\bar{u}(x)$ in
Refs.~\cite{zzy02,ah05} and $\bar{d}(x)/\bar{u}(x)$ in
Ref.~\cite{ah05} have similar behavior as ours.

\begin{figure}
\begin{center}
\includegraphics[scale=1]{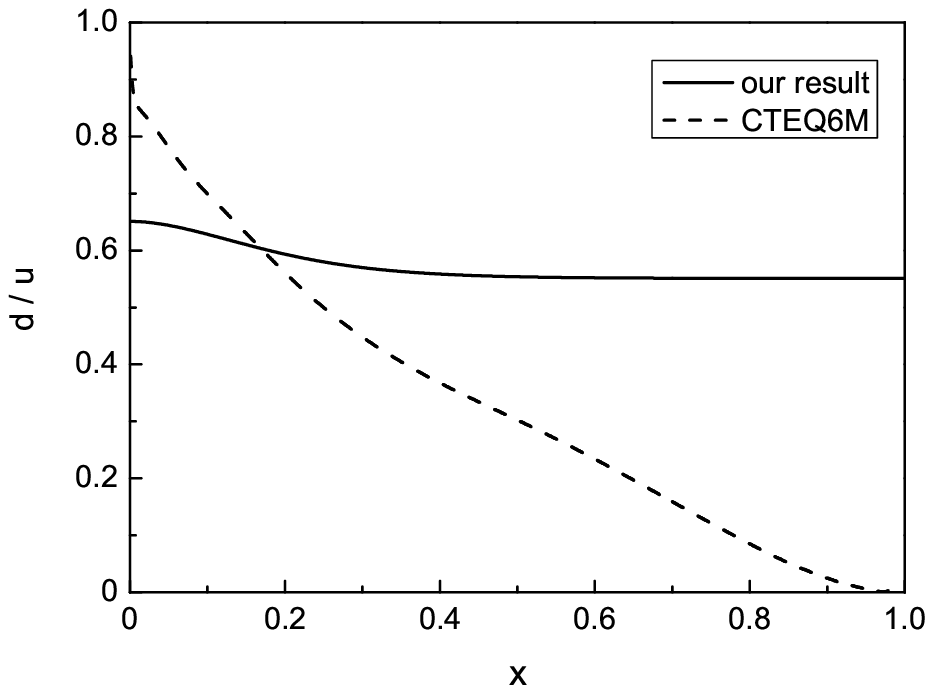}
\caption{Comparison of our result with CTEQ result at $Q^2=1$
GeV$^2$~\cite{cteq02} for $d(x)/u(x)$.}\label{dou}
\end{center}
\end{figure}

Furthermore, $d(x)/u(x)$ is also shown in Fig.~\ref{dou}. Note that
when $x\rightarrow 1$, $d(x)/u(x)\rightarrow0.55$, which is
different from the result of CTEQ parametrization~\cite{cteq02}, but
close to the prediction of the naive SU(6) quark model.

\begin{figure}
\begin{center}
\includegraphics[scale=1]{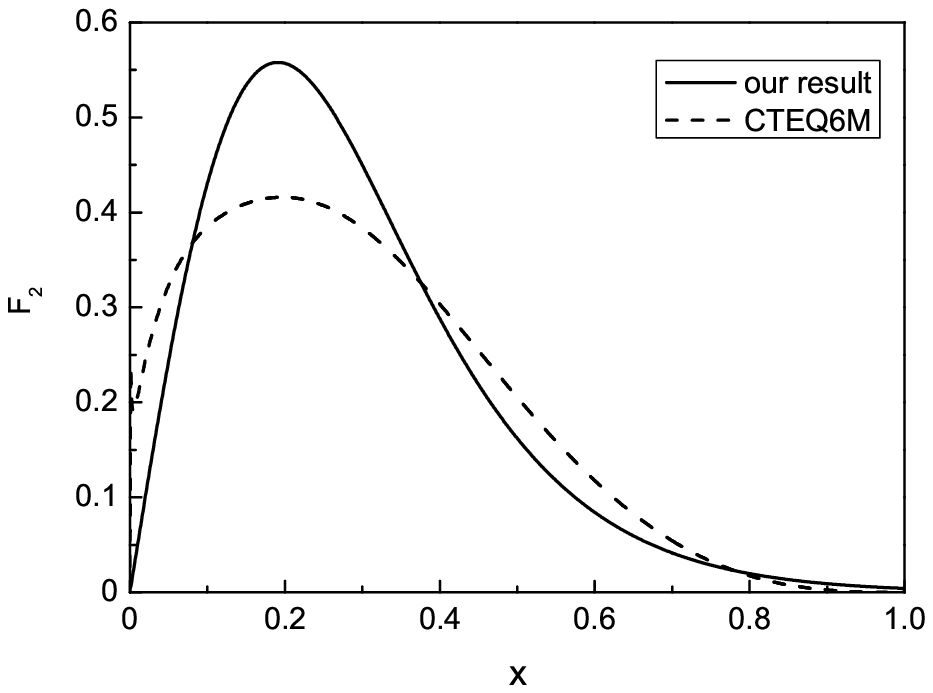}
\caption{Comparison of our result with CTEQ result at $Q^2=1$
GeV$^2$~\cite{cteq02} for $F_2(x)$.}\label{f2}
\end{center}
\end{figure}

\begin{figure}
\begin{center}
\includegraphics[scale=1]{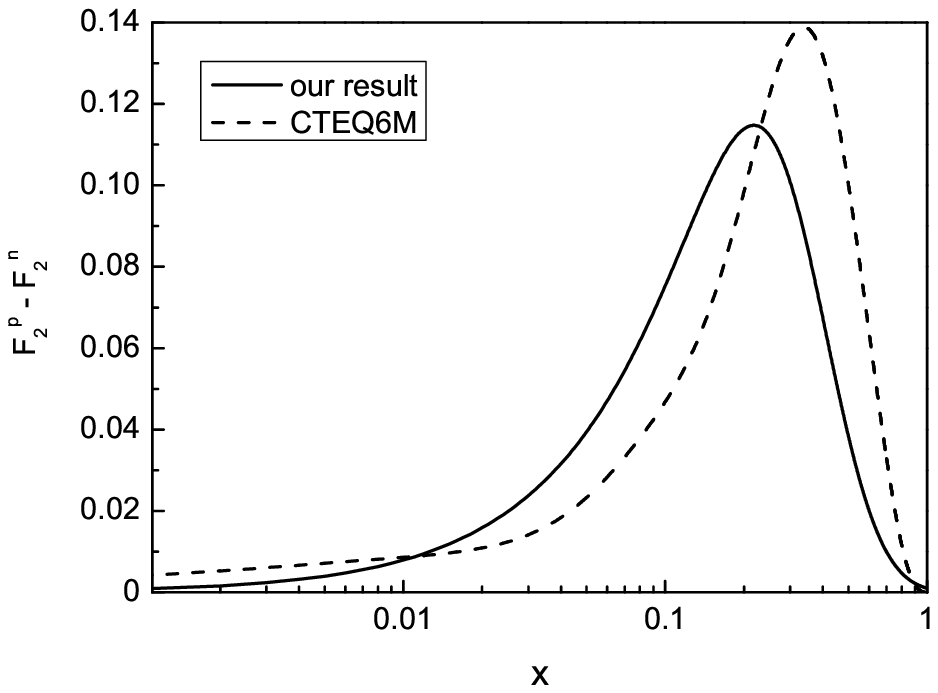}
\caption{Comparison of our result with CTEQ result at $Q^2=1$
GeV$^2$~\cite{cteq02} for $F_2^p(x)-F_2^n(x)$.}\label{f2mf2n}
\end{center}
\end{figure}

\begin{figure}
\begin{center}
\includegraphics[scale=1]{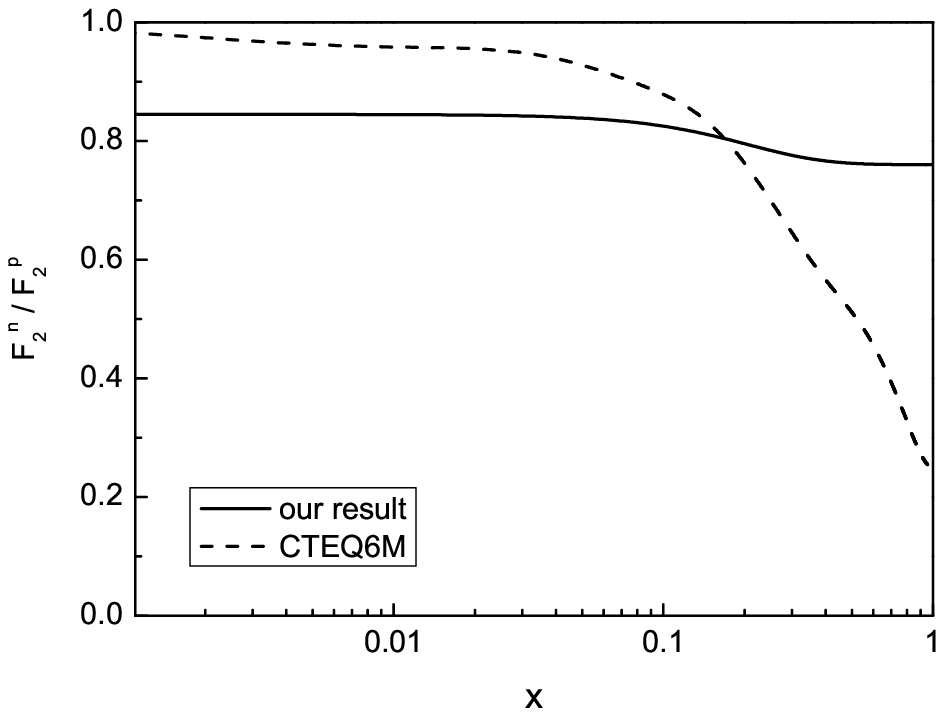}
\caption{Comparison of our result with CTEQ result at $Q^2=1$
GeV$^2$~\cite{cteq02} for $F_2^p(x)/F_2^n(x)$.}\label{f2nof2}
\end{center}
\end{figure}

The nucleon structure function $F_2(x)=2xF_1(x)=x\sum_fe_f^2f(x)$,
where $e_f$ is the charge of the parton of flavor $f$, is shown in
Fig.~\ref{f2}. With the $p$-$n$ isospin symmetry, i.e. $u^n(x,{\bf
k}_\bot)=d^p(x,{\bf k}_\bot)$, $d^n(x,{\bf k}_\bot)=u^p(x,{\bf
k}_\bot)$, $\bar{u}^n(x,{\bf k}_\bot)=\bar{d}^p(x,{\bf k}_\bot)$,
$\bar{d}^n(x,{\bf k}_\bot)=\bar{u}^p(x,{\bf k}_\bot)$, $g^n(x,{\bf
k}_\bot)=g^p(x,{\bf k}_\bot)$, we can also obtain the structure
function of the neutron $F_2^n(x)$. $F_2^p(x)-F_2^n(x)$ and
$F_2^n(x)/F_2^p(x)$ are shown in Fig.~\ref{f2mf2n} and
Fig.~\ref{f2nof2}. We can see that $F_2^p(x)-F_2^n(x)$ is quite
agreeable, while $F_2^n(x)/F_2^p(x)$ is not. One interesting feature
is the behavior of $F_2^n(x)/F_2^p(x)$ when $x\rightarrow 1$. In the
naive SU(6) quark model it tends to $2/3$, while in the SU(6)
quark-diquark model it is $1/4$ and in experimental observation it
is smaller than $1/2$. Here our result seems close to the prediction
of the naive SU(6) quark model again, however, it does not agree
with the reality.

\begin{figure}
\begin{center}
\includegraphics[scale=1]{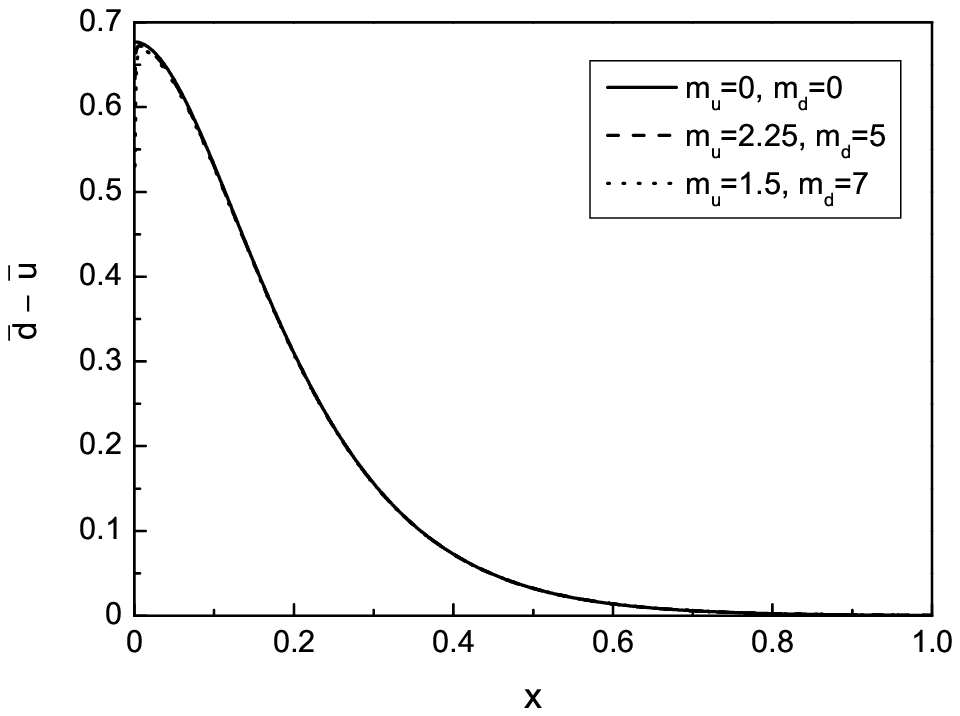}
\includegraphics[scale=1]{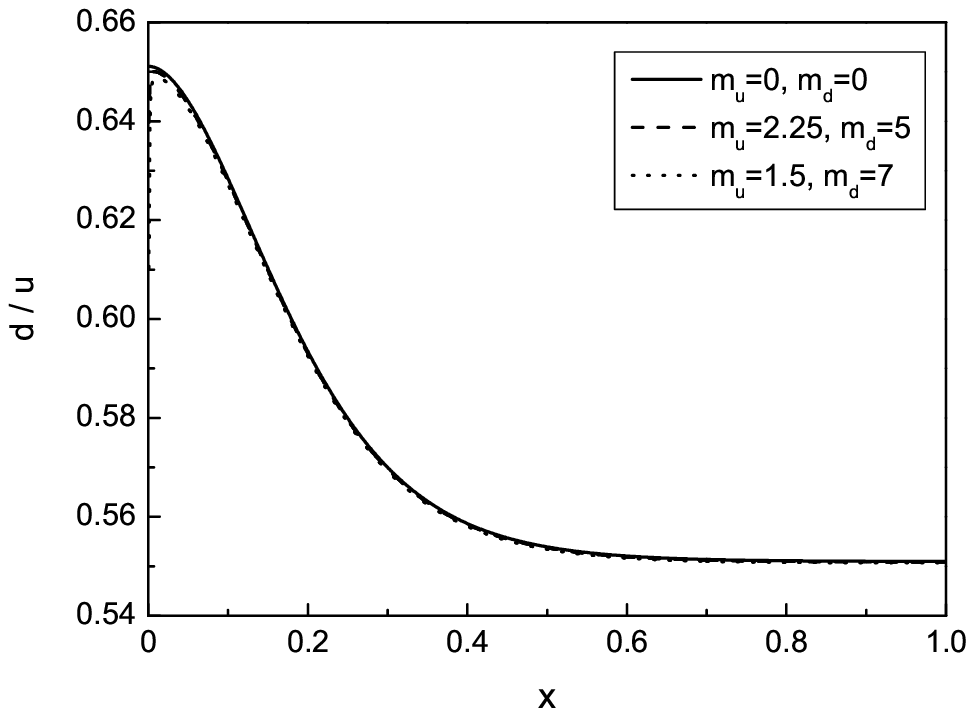}
\caption{$\bar{d}(x)-\bar{u}(x)$ and $d(x)/u(x)$ at different masses
(unit: MeV) of $u$, $d$ quark.}\label{udmass}
\end{center}
\end{figure}

From the above results, we find that our statistical method can
successfully describe the behavior of the ``subtracted--terms'',
such as $\bar{d}(x)-\bar{u}(x)$ and $F_2^p(x)-F_2^n(x)$, but the
``divided--terms'', such as $\bar{d}(x)/\bar{u}(x)$, $d(x)/u(x)$ and
$F_2^p(x)/F_2^n(x)$, can only match the trend of experimental
results approximately, and the departure is especially large in the
high-$x$ region, where the valence parts of the PDFs dominate. This
feature probably implies that an additive statistics-irrelevant
corrected term to the PDFs works, whereas more free parameters and
uncertainty should be introduced. Bhalerao {\it et al.} successfully
reproduced most features of the PDFs and structure functions by
introducing two extra corrected terms~\cite{bh95,bh96,bkr00,bh00},
at the cost of two more free parameters and violating the $p$-$n$
isospin symmetry.

\section{Further discussions}
We have ignored the masses of the quarks and anti-quarks for
simplicity. Nevertheless, mass effect can be taken into account
without difficulty. Actually we have performed this calculation and
found, as can be speculated, the correction of mass effect to the
light-flavored PDFs is very small. $\bar{d}(x)-\bar{u}(x)$ and
$d(x)/u(x)$ with different masses of $u$, $d$ quark are illustrated
in Fig.~\ref{udmass}.

However, the mass difference between $u$ and $d$ quarks can generate
the mass split between the proton and the neutron. The invariant
mass square of the system is given by
\begin{equation}
\label{mm} M^2 = \sum_i\left(\frac{m^2+k_\bot^2}{x}\right)_i\;.
\end{equation}

In the continuous condition, it is

\begin{equation}
\label{mmc}
M^2=\sum_f\int\left[\iint\left(\frac{m_f^2+k_\bot^2}{x}\right)f(x,k_\bot)\,\mathrm{d}^2k_\bot\right]\mathrm{d}x\;.
\end{equation}

Using the $p$-$n$ isospin symmetry, we get
\begin{equation}
\label{dm}
M_n^2-M_p^2=(m_d^2-m_u^2)\int\frac{1}{x}\left[u^p(x)+\bar{u}^p(x)-d^p(x)-\bar{d}^p(x)\right]\mathrm{d}x\;.
\end{equation}

In PDG 2006, $m_u=1.5\sim 4$ MeV, $m_d=3\sim 7$ MeV,
$\Delta_{pn}=M_n-M_p=939.565-938.272=1.293$ MeV.

When we use the mean value $m_u=2.25$ MeV, $m_d=5$ MeV, we get
$\Delta_{pn}=0.664$ MeV from Eq.~(\ref{dm}), and when we use the
extreme value $m_u=1.5$ MeV, $m_d=7$ MeV, then $\Delta_{pn}=1.557$
MeV. The result seems rather agreeable and it reinforces the
reasonableness of our approach.

We have only calculated the light-flavored PDFs, however, the heavy
flavors, such as $s$, $c$, $b$, $t$ quarks and the corresponding
anti-quarks, can be treated in the same way. Since, in nucleon, the
valence numbers of them are zero, the chemical potentials of them
must all be zero. It leads directly to three following conclusions:

Firstly, except for their masses, the heavy-flavored PDFs have no
additional free parameters than the light-flavored ones. That is, if
their masses are used as inputs, their PDFs can be uniquely
determined by the parameters $T$ and $V$, which have already been
fixed in the previous light-flavored condition. So the difference
between heavy-flavored PDFs only comes from the difference of their
masses.

Secondly, the quark and anti-quark of the same heavy flavor have
just the same distribution. For example, the condition
$s(x)=\bar{s}(x)$ holds in the whole region $x\in[0,1]$, so that the
$s$, $\bar{s}$ asymmetry in the nucleon~\cite{bmdxm} does not come
from the pure statistical effect.

\begin{figure}
\begin{center}
\includegraphics[scale=1]{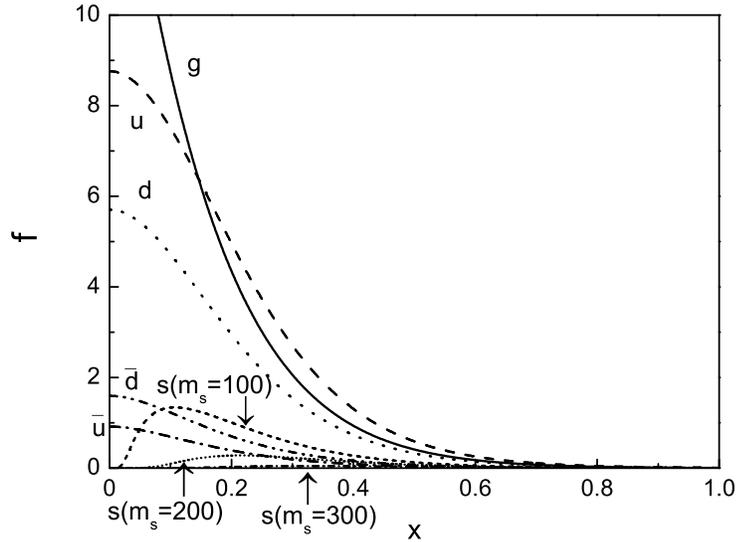}
\caption{Comparison of $s(x)$ at different mass ($m_s=100, 200, 300$
MeV) with the light-flavor PDFs at $T_0$ and $V_0$.}\label{udgs}
\end{center}
\end{figure}

Thirdly, from Eq.~(\ref{fx}), we can see that $f(x)$ decreases when
$m_f$ increases. $s(x)$ with different $m_s$ at $T_0$ and $V_0$,
together with the light-flavored PDFs, are illustrated in
Fig.~\ref{udgs}, and it shows that when $m_s \leq 100$ MeV the
contribution of $s$ quark is considerable, and when $m_s>200$ MeV it
is minor. Therefore, the contribution of heavier flavors is
negligible. Calculation also indicates that $s$ quark contributes
less than $7\%$ both to the total light-front momentum fraction $x$
(see Eq.~(\ref{qnor})) and to the total invariant mass square of the
system at $T_0$ and $V_0$ (see Eq.~(\ref{mmc})).

\section{Summary}
We preform a simple statistical approach and obtain analytic
expressions of the parton distribution functions in terms of
light-front kinematic variables in the whole $x$ region [0,1]. The
low-$x$ behavior of these parton distribution functions is different
from those in some previous instant-form statistical models in the
infinite-momentum frame and our results are more close to the
reality. There is no arbitrary parameter or extra corrected term put
by hand in our model, which guarantees the stringency of our
conclusion. Several features of the parton distribution functions
and structure functions of the nucleon are compared with the results
of experiments and other theories. Calculation of the mass split
between the proton and the neutron is also performed. We have
further discussions on the influence of the heavy flavors. All of
these show that although the statistical effect is not everything,
it is very important to some aspects of the nucleon structure.

\section*{Acknowledgment}

This work is partially supported by National Natural Science
Foundation of China (Nos.~10721063, 10575003, 10528510), by the Key
Grant Project of Chinese Ministry of Education (No.~305001), by the
Research Fund for the Doctoral Program of Higher Education (China).
It is also supported by Hui-Chun Chin and Tsung-Dao Lee Chinese
Undergraduate Research Endowment (Chun-Tsung Endowment) at Peking
University, and by National Fund for Fostering Talents of Basic
Science (No.~J0730316 and No.~J0630311).


\begin{thebibliography}{99}

\bibitem {f69} R.P. Feynman,
Proceedings of the 3rd Topical Conference on High Energy Collision
of Hadrons, Stony Brook, N. Y. (1969).

\bibitem {bp69} J.D. Bjorken and E.A. Paschos,
\Journal{\PRV}{185}{1975}{1969}.

\bibitem {c87} J. Cleymans,
presented at Int. Workshop on Quarks, Gluons and Hadronic Matter,
Cape Town, South Africa, Feb 16-20, 1987. Published in Cape Town
Workshop 1987:0447 (QCD161:C375:1987).

\bibitem {ct88} J. Cleymans and R.L. Thews,
\Journal{\ZPC}{37}{315}{1988}.

\bibitem {mu89} E. Mac and E. Ugaz,
\Journal{\ZPC}{43}{655}{1989}.

\bibitem {bl90} R.P. Bickerstaff and J.T. Londergan,
\Journal{\PRD}{42}{3621}{1990}.

\bibitem {cdj93} J. Cleymans, I. Dadic, and J. Joubert,
 arXiv:hep-ph/9307352.

\bibitem {gdr91} K. Ganesamurthy, V. Devanathan, and M. Rajasekaran,
\Journal{\ZPC}{52}{589}{1991}.

\bibitem {dkg94} V. Devanathan, S. Karthiyayini, and K. Ganesamurthy,
\Journal{\MPLA}{9}{3455}{1994}.

\bibitem {dm96} V. Devanathan and J.S. McCarthy,
\Journal{\MPLA}{11}{147}{1996}.

\bibitem {bh95} R.S. Bhalerao,
 arXiv:hep-ph/9506367.

\bibitem {bh96} R.S. Bhalerao,
\Journal{\PLB}{380}{1}{1996}.

\bibitem {bkr00} R.S. Bhalerao, N.G. Kelkar, and B. Ram,
\Journal{\PLB}{476}{285}{2000}.

\bibitem {bh00} R.S. Bhalerao,
 arXiv:hep-ph/0003075.

\bibitem {bs95} C. Bourrely and J. Soffer,
\Journal{\NPB}{445}{341}{1995}.

\bibitem {bsb02} C. Bourrely, J. Soffer, and F. Buccella,
\Journal{\EPJC}{23}{487}{2002}.

\bibitem {bsb05} C. Bourrely, J. Soffer, and F. Buccella,
\Journal{\EPJC}{41}{327}{2005}.

\bibitem {zzm01} Y.-J. Zhang, B. Zhang, and B.-Q. Ma,
\Journal{\PLB}{523}{260}{2001}.

\bibitem {zzy02} Y.-J. Zhang, B.-S. Zou, and L.-M. Yang,
\Journal{\PLB}{528}{228}{2002}.

\bibitem {zdm02} Y.-J. Zhang, W.-Z. Deng, and B.-Q. Ma,
\Journal{\PRD}{65}{114005}{2002}.

\bibitem {ah05} M. Alberg and E.M. Henley,
\Journal{\PLB}{611}{111}{2005}.

\bibitem {bpp98} For a review, see, {\it e.g.}, S.J. Brodsky, H.C. Pauli, and S.S. Pinsky,
\Journal{\PR}{301}{299}{1998}.

\bibitem {ms90} B.-Q. Ma and J. Sun,
\Journal{\IJMPA}{6}{345}{1991}; B.-Q. Ma and J. Sun,
\Journal{\JPG}{16}{823}{1990}.

\bibitem {adp02} V.S. Alves, A. Das, and S. Perez,
\Journal{\PRD}{66}{125008}{2002}.

\bibitem {w03} H.A. Weldon,
\Journal{\PRD}{67}{085027}{2003}.

\bibitem {bmfw03} M. Beyer, S. Mattiello, T. Frederico, and H.J. Weber,
 arXiv:hep-ph/0310222.

\bibitem {rb04} J. Raufeisen and S.J. Brodsky,
\Journal{\PRD}{70}{085017}{2004}.

\bibitem {e1907} A. Einstein,
\Journal{\JRE}{4}{411}{1907}.

\bibitem {p58} W. Pauli, Theory of relativity, (Pergamon Press,
Oxford, United Kingdom, 1958).

\bibitem {o63} H. Ott,
\Journal{\ZP}{175}{70}{1963}.

\bibitem {hw71} D. Ter Haar and H. Wergeland,
\Journal{\PR}{1}{31}{1971}.

\bibitem {cm95} S.S. Costa and G.E.A. Matsas,
\Journal{\PLA}{209}{155}{1995}.

\bibitem {lm04} P.T. Landsberg and G.E.A. Matsas,
\Journal{\PA}{340}{92}{2004}.

\bibitem {r0801} M. Requardt,
 arXiv:0801.2639 [gr-qc].

\bibitem {cpdth07} D. Cubero, J. Casado-Pascual, J. Dunkel, P. Talkner, and P. H\"{a}nggi,
\Journal{\PRL}{99}{170601}{2007}.

\bibitem {r0804} C. Rasinariu,
 arXiv:0804.3836 [gr-qc].

\bibitem {nmc9194} New Muon Collaboration, P. Amaudruz {\it et al.},
\Journal{\PRL}{66}{2712}{1991}; M. Arneodo {\it et al.},
\Journal{\PRD}{50}{R1}{1994}.

\bibitem {cteq02} J. Pumplin {\it et al.},
\Journal{\JHEP}{0207}{012}{2002}.

\bibitem {e866989901} FNAL E866/NuSea Collaboration, E.A. Hawker {\it et
al.}, \Journal{\PRL}{80}{3715}{1998}; J.C. Peng {\it et al.},
\Journal{\PRD}{58}{092004}{1998};
 R.S. Towell, P.L.
McGaughey {\it et al.}, \Journal{\PRD}{64}{052002}{2001}.

\bibitem {na94} NA51 Collaboration, A. Baldit {\it et
al.}, \Journal{\PLB}{332}{244}{1994}.

\bibitem {hermes98} HERMES Collaboration, K. Ackerstaff {\it et
al.}, \Journal{\PRL}{81}{5519}{1998}.

\bibitem {bmdxm} See, {\it e.g.}, S.J. Brodsky and B.-Q. Ma, \Journal{\PLB}{381}{317}{1996};
Y. Ding and B.-Q. Ma, \Journal{\PLB}{590}{216}{2004}; Y. Ding, R.-G.
Xu, and B.-Q. Ma, \Journal{\PLB}{607}{101}{2005};
\Journal{\PRD}{71}{094014}{2005}.

\end{thebibliography}
\end{document}